\documentclass[12pt]{article}
\usepackage[dvips]{graphicx}
\usepackage{epsfig,amsmath}
\begin{document}
\thispagestyle{empty}
\begin{center}

{\bf DSPIN-09 WORKSHOP SUMMARY \footnote{Invited talk presented at the Workshop DUBNA-SPIN 09, Sept. 01-05, 2009, Dubna (Russia), to appear in the Proceedings.}}

\vskip 1.40cm
{\bf Jacques Soffer}
\vskip 0.3cm
{\it Physics Department, Temple University\\
Barton Hall, 1900 N, 13th Street\\
Philadelphia, PA 19122-6082, USA} 
\vskip 1.0cm
{\bf Abstract}
\end{center}

I will try to summarize several stimulating open questions in high energy spin physics, which were discussed during the five
days of this workshop, showing also the striking progress recently achieved in this field.

\vspace{7.2mm} 

\section{Introduction}
Once again this workshop has been very productive with a high density scientific program, since about 95 talks
were presented. It has facilitated detailed discussions between theorists and experimentalists and, moreover since spin occurs
in all particle  processes, it is obvious that by ignoring this fundamental tool, we will miss an important part of the story. I will mention some substantial progress which have been achieved since DSPIN-07 and, whenever it is possible, to identify what we have learnt and what are the prospects. Although I will not cover technical subjects, I just want to mention some future projects, in particular E. Steffens, who reported about the status of the challenging PAX experiment in FAIR at GSI, supplemented by a bright theorist perspective by N. Nikolaev on polarized antiproton experiments. S. Nurushev gave a talk on the polarization program SPASCHARM at IHEP, in connection with the latest results from the PROZA experiment and various aspects of the preparation of the spin program at the Nuclotron in Dubna, were presented by Y. Gurchin, S. Piyadin and P. Kurilkin.\\ 
The opening talk of DSPIN-09 was given by A. Krisch who recalled us some unexpected large spin effects in $pp$ elastic scattering obtained about twenty years ago, whose clear theoretical interpretation is not yet available (see Figs.~1a,b). These results provided the motivation to undertake a 
very successful Siberian Snake program, allowing to obtain high energy polarized proton beams, an essential element of the RHIC Spin program at BNL. Following the introduction, this summary talk contains several sections, first on the experimental side, with new results from COMPASS, HERMES, BELLE, JLab and RHIC and at the end a section devoted to several theoretical subjects.
\section{New results from COMPASS}
The COMPASS fixed-target experiment at the CERN SPS has produced over the last two years several interesting new results in different areas. R. Gazda presented an analysis of 2002-2007 data on the first moment of the structure function $g_1$ and semi-inclusive asymmetries, on proton and deuteron targets, which has led to a better determination of the polarized valence quark and sea quark distributions with flavor separation. Concerning the
important issue of the gluon polarization it has been extracted from high $p_T$ hadron pairs, either in the quasi-real photoproduction ($Q^2<1\mbox{GeV}^2$) or in the DIS ($Q^2>1\mbox{GeV}^2$) regimes. L. Silva concluded that two independent analyses lead to compatible values, with
a result consistent with zero, as shown in Fig.~2. One can also see in Fig.~2 that this small gluon polarization was confirmed by the method based on the measurement of open-charm asymmetries, as reported by K. Kurek, supplemented by a NLO QCD prediction.\\
\begin{figure}[h]
  \centering
  \begin{tabular}{cc}
  
    \includegraphics[width=45mm]{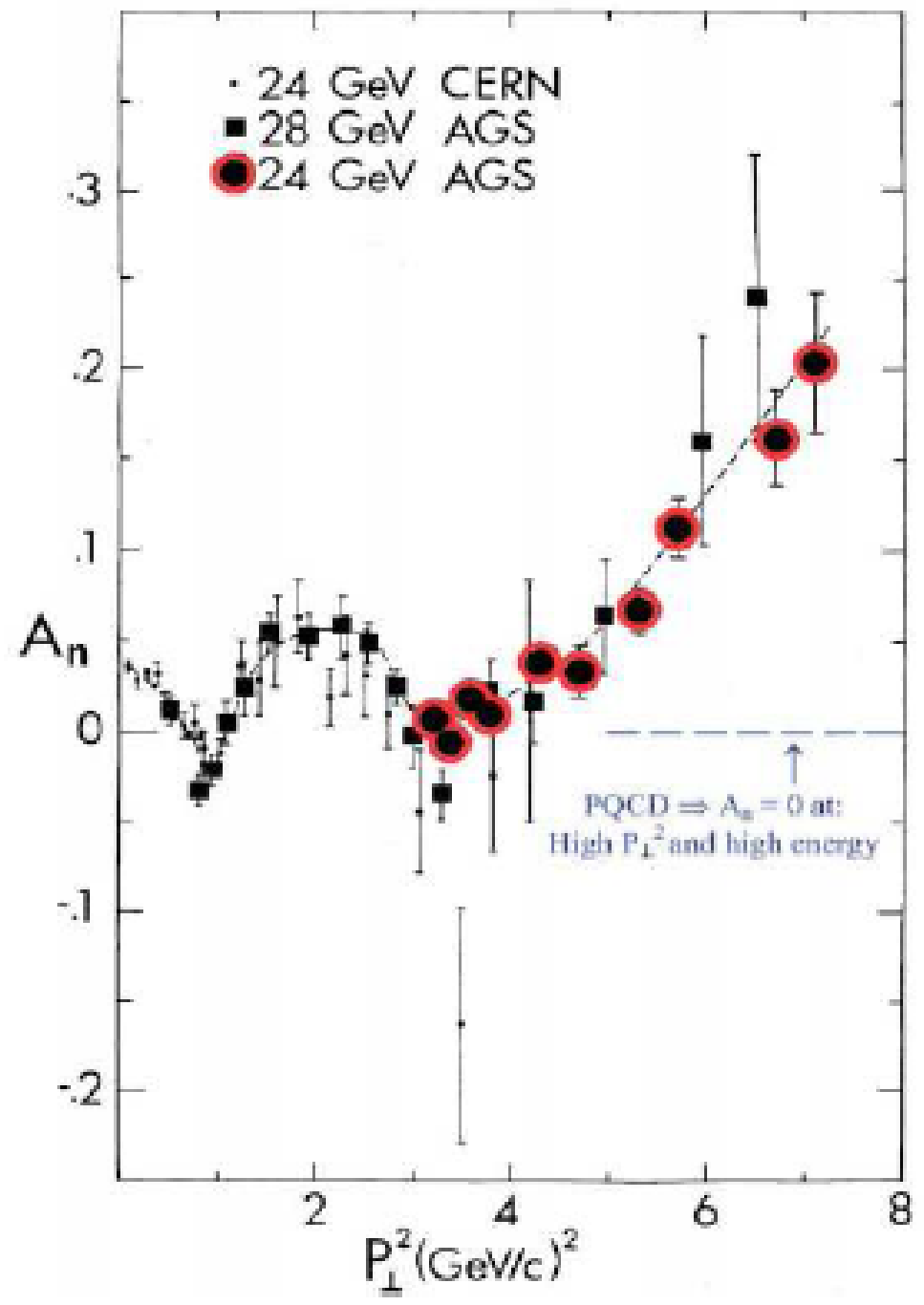} &
    \includegraphics[width=65mm]{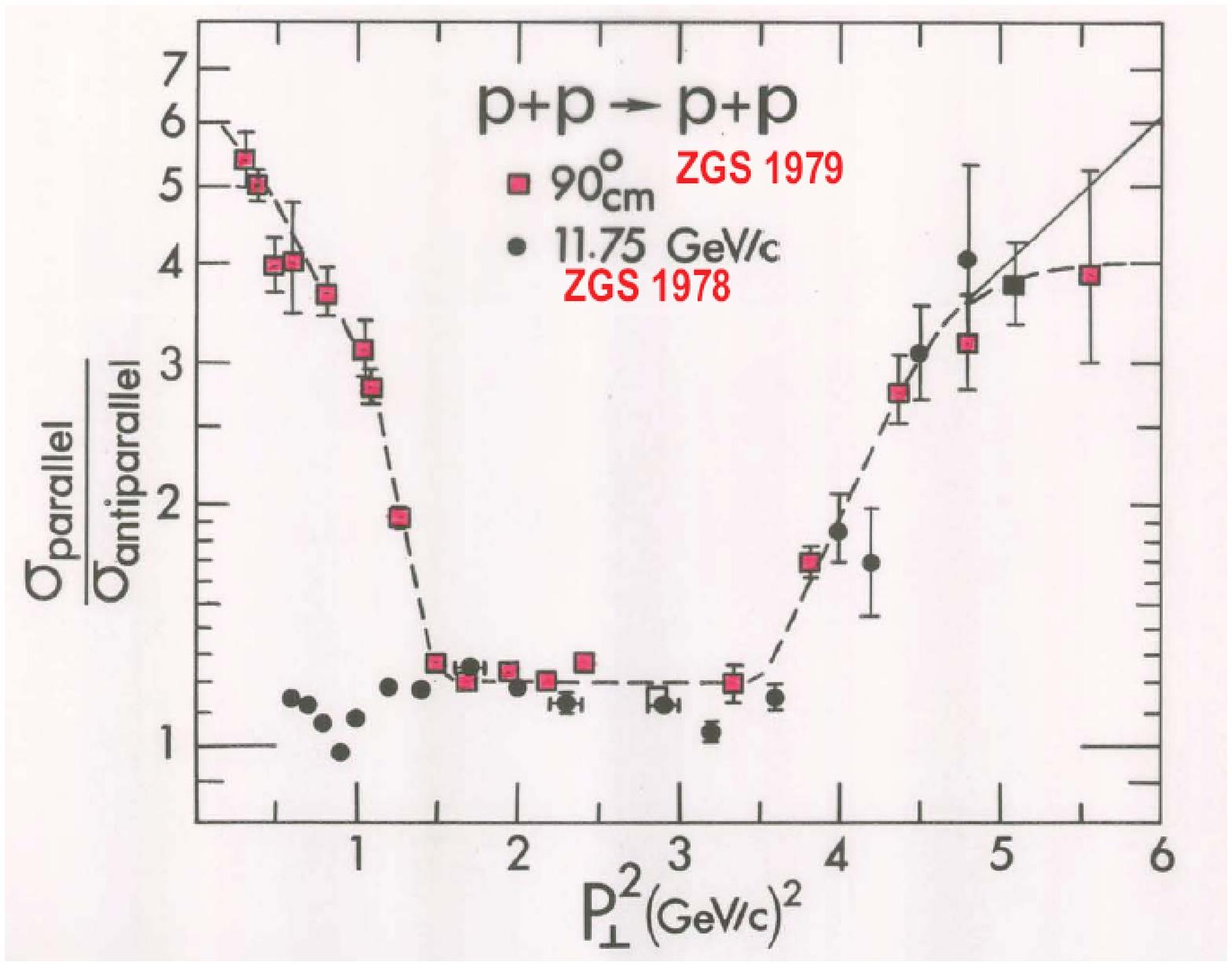}\\
    \textbf{(a)} & \textbf{(b)}
  \end{tabular}
  \caption{%
    \textbf{(a)} The single-spin asymmetry $A_n$ in $pp$ elastic scattering as a function of $p_T^2$
    \textbf{(b)} Ratio of $pp$ differential cross sections for spins parallel or antiparallel along the normal to the scattering plane ( Both taken from Krisch's talk \cite{ADK}).
   }
  \label{yourname_fig2}
\end{figure}
%%%%%%%%%%%%%%%%%%%%%%%
%\begin{center}
\begin{figure}[h]
\hspace{6pc}
\includegraphics[width=25pc]{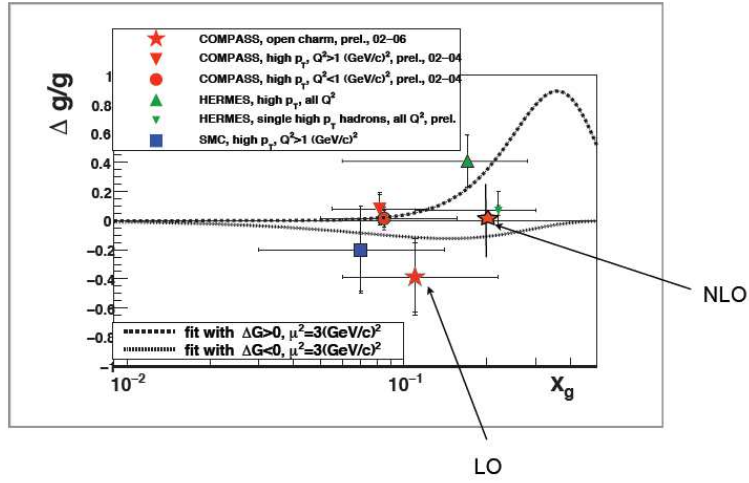}
\caption{\label{label}
Comparison of the $\Delta G/G$ measurements from various experiments (Taken from Kurek's talk \cite{KK}).
}
\end{figure}
%\end{center}
Longitudinal polarization of the $\Lambda$ and $\bar \Lambda$  hyperons DIS events collected by COMPASS in 2003-2004, the world largest sample, was presented by V. Rapatskiy, with the goal to study $\Lambda$ spin structure models. The results will be improved by adding a much larger events number from 2006-2007 data. The measurements of azimuthal asymmetries in semi-inclusive hadron production with a longitudinally polarized deuterium target, led to some preliminary results reported by I. Savin, in various kinematical variables. The COMPASS experiment has put a serious effort in the determination of the transverse momentum dependent parton distributions in particular the Collins function, which gives access to the transversity
distribution, and the Sivers function. Several new results were given in G. Sbrizzai's talk as shown in Figs.~3a,b and although the Collins asymmetries
on proton are compatible with the predictions of the present picture, the agreement is marginal for the Sivers asymmetries. New data on transversely
polarized proton will be taken in 2010 and the large increase in precision should, hopefully, solve this important problem.\\
The Generalized Parton Distributions (GPD) program proposed by COMPASS was covered by A. Sandacz. This project will explore intermediate $x$ (0.01 - 0.1) and large $Q^2$ (8 - 12)$\mbox{GeV}^2$ and will be unique in this kinematical range before availability of new colliders. There is another future project by COMPASS, a complete program of Drell-Yan experiments for probing the hadron structure, with first data taking after 2012, which was presented by O. Denisov. Drell-Yan process is very rich because unpolarized angular distributions give access to the Boer-Mulders function, single spin-asymmetry allows to test the sign change from SIDIS of the Sivers function, a fundamental test of gauge theory and the double transverse spin asymmetry gives access to the transversity distributions. 
\begin{figure}[h]
  \centering
  \begin{tabular}{cc}
  \hspace{-2pc}
    \includegraphics[width=85mm]{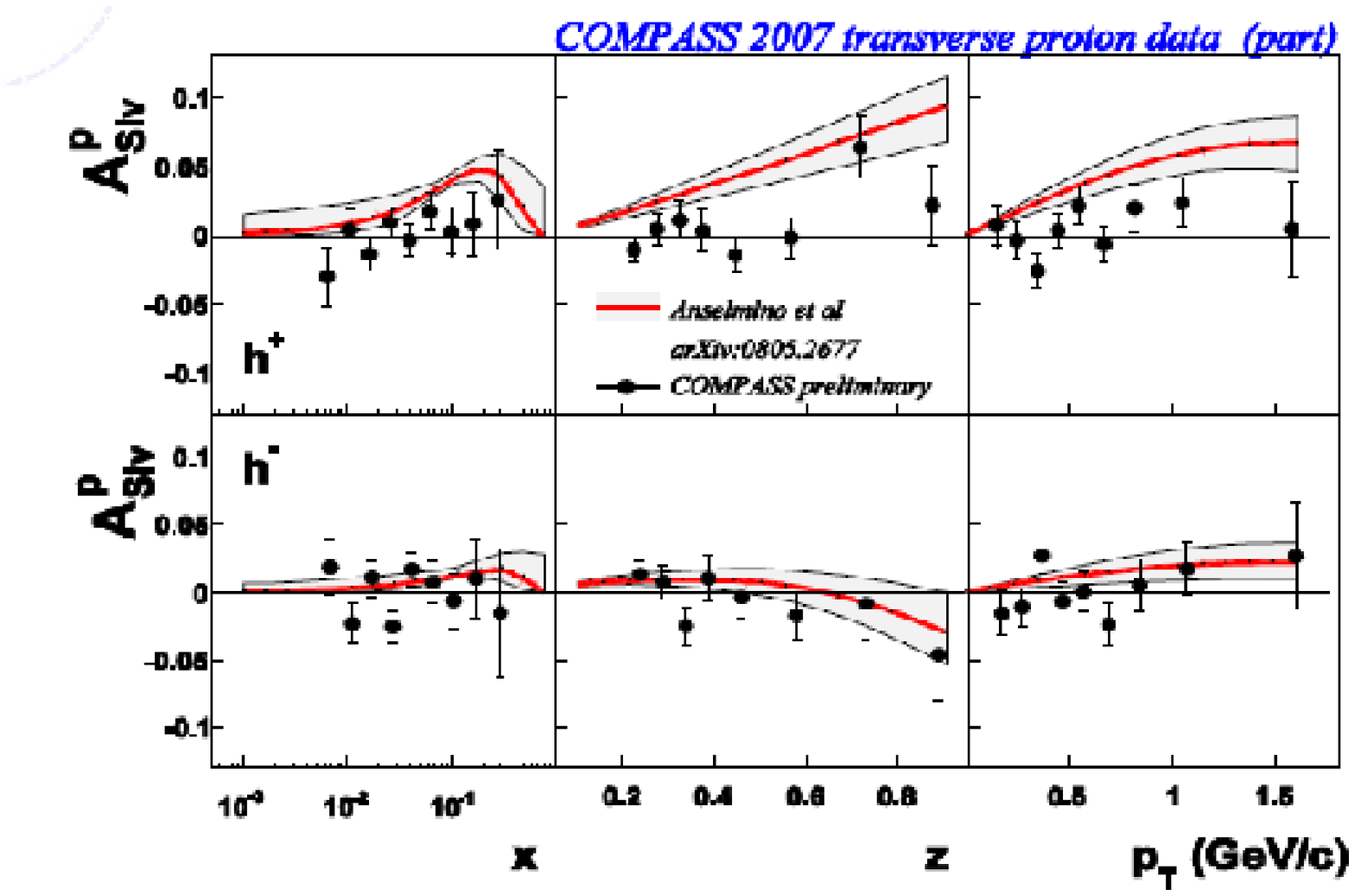} &
    \includegraphics[width=80mm]{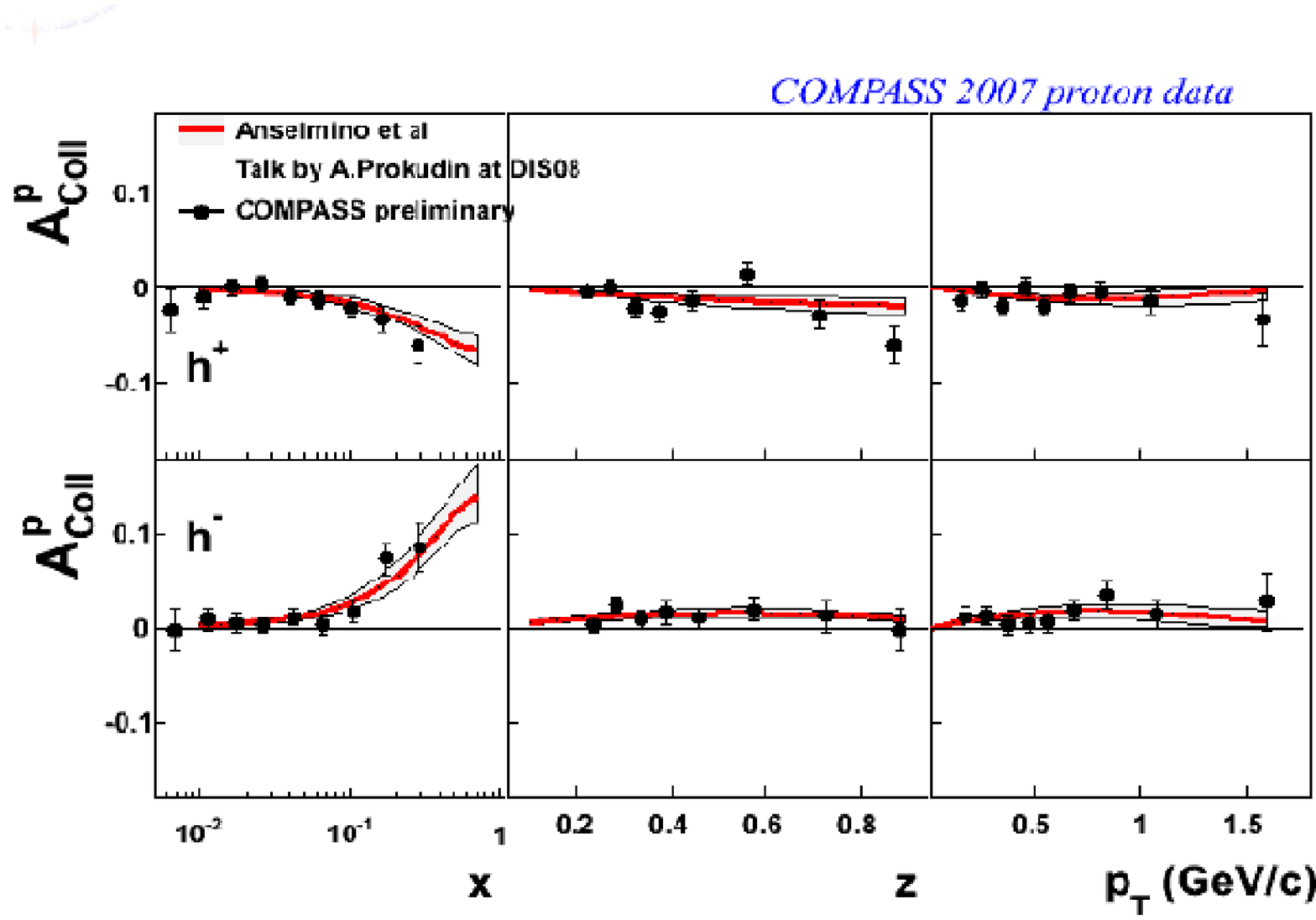}\\ 
    \textbf{(a)} & \textbf{(b)}
  \end{tabular}
  \caption{%
    \textbf{(a)} COMPASS Sivers asymmetries on proton versus $x$, $z$ and $p_T$.
    \textbf{(b)} COMPASS Collins asymmetries on proton versus $x$, $z$ and $p_T$ (Both taken from Sbrizzai's talk \cite{GS}).
   }
  \label{yourname_fig2}
\end{figure}
\section{New results from HERMES}
An overview of the HERMES experiment was given by V. Korotkov, including in the pure DIS sector, latest results on the
$F_2^{p,d}$ and $g_2^p$ structure functions and in the SIDIS sector, Collins asymmetries, Sivers asymmetries, azimuthal asymmetries in unpolarized scattering and strange quark distributions.\\
New results on Deeply Virtual Compton Scattering (DVCS) at HERMES were reported by A. Borissov, whose motivation is to get access to GPD. DVCS azimuthal asymmetries on proton provide a constraint on total angular momentum of valence quarks and allow the comparison with GPD model calculations. Along the same lines, S. Manayenkov gave a talk on a detailed study of exclusive electroproduction of $\rho^0$, $\phi$ and $\omega$ at HERMES, with a special emphasis on the extraction of the spin density matrix elements and tests of s-channel helicity conservation (SCHC). Violation of SCHC is observed in $\rho^0$ production, both on proton and deuteron, but no such signal is found for $\phi$ meson. Finally $\Lambda$ physics at HERMES was
covered by Y. Naryshkin, where they observed for $\Lambda$ and $\bar \Lambda$, the longitudinal spin transfer from the beam, shown in Fig.~4, a very small longitudinal spin transfer from the target and a reliable transverse polarization.
\begin{figure}[h]
\hspace{4pc}
\includegraphics[width=28pc]{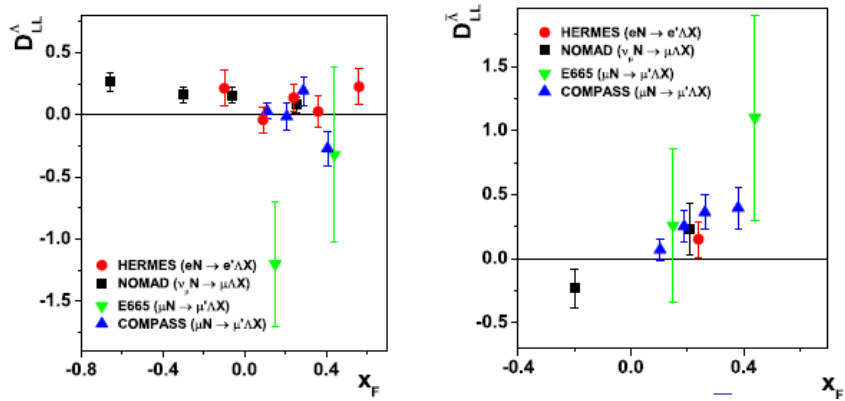}
\caption{\label{label}
Spin transfer parameter $D_{LL}$ versus $x_F$ for $\Lambda$ and $\bar \Lambda$ (Taken from Naryshkin's talk \cite{YN}).}
\end{figure}
%\clearpage
\section{New results from BELLE and JLab}
The measurement of quark transversity distributions can be done by different methods, using either $pp$ collisions or SIDIS and $e^+e^-$ collisions. A. Vossen recalled the first attempt to extract them, in a model dependent way, from BELLE data combined with Collins asymmetry in SIDIS. Then he described the forthcoming observation of an interference fragmentation function asymmetry in $e^+e^-$ collisions at BELLE, another very promising method.\\
An introductory review talk of spin physics with CLAS in Hall B at JLab, was presented by Y. Prok, in particular some recent experimental results. The list of topics was covering, the status of $g_1^p(x,Q^2)$, including the resonance region, the large-$x$ behavior of the $A_1$ asymmetry, dominated by valence quarks, the generalized Gerasimov-Drell-Hearn (GDH) sum rule, semi-inclusive processes, etc...The whole activity with CLAS was
supplemented by several presentations. First, M. Mirazita with studies in SIDIS with longitudinal polarization and future transversely polarized target, in view of the determination of the transverse momentum dependent distributions. Second, C. Munos Camacho with a presentation of the GPD experimental program at JLab, measurements of DVCS/GPD both in Hall B and in Hall A, together with some exciting perspectives for the future with the 12GeV upgrade. Third, V. Drozdov told us that, concerning the CLAS EG4 experiment on the GDH sum rule in the low $Q^2$ region, the analysis is underway. Finally, E. Pasyuk who reported on very interesting results from CLAS on several polarization measurements, in exclusive photoproduction, after the new addition of a frozen spin target, with both longitudinal and transverse polarization.\\
A new measurement of the polarization transfer in $ep$ elastic scattering, by the recoil polarization technique, was done in Hall C at JLab, allowing to extract the ratio of electric and magnetic form factors $G_{Ep}/G_{Mp}$ at large $Q^2$. The preliminary results were presented by V. Punjabi and are shown, with earlier results, in Fig.~5. They confirm the previous observation that this ratio decreases with increasing $Q^2$ and contradicts the expectation of a flat behavior, according to conventional wisdom. This measurement will be done up to $Q^2= 15\mbox{GeV}^2$ with the 12GeV upgrade. Finally, the search for a 2$\gamma$ contribution in $ep$ elastic scattering, a very closely related subject, was presented by Ch. Perdrisat, together with some preliminary results of the $G_{Ep}(2\gamma)$ experiment at JLab.

\begin{figure}[h]
\hspace{6pc} 
\includegraphics[width=22pc]{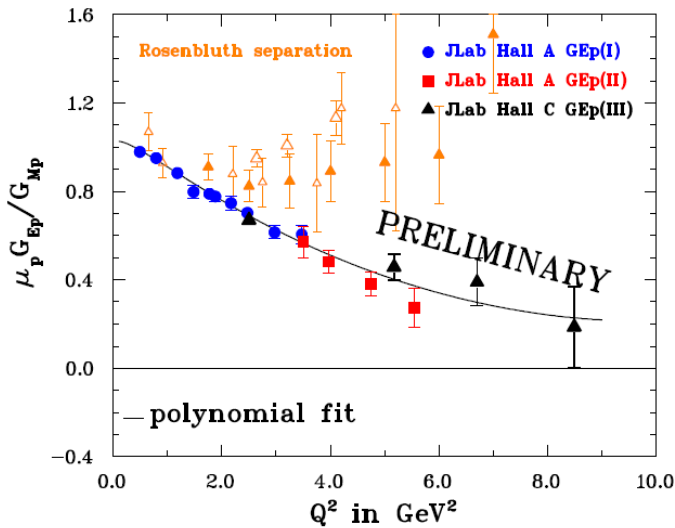}
\caption{\label{label}
The proton elastic form factor ratio versus
$Q^2$ (Taken from Punjabi's talk \cite{VP}).}
\end{figure}

\section{New results from RHIC}
Brookhaven National Lab. operates since 2001 a polarized $pp$ collider in RHIC, which will be running up to $\sqrt{s}$=500GeV, to perform a vast
program of spin measurements. The energy is high enough to assume that NLO pQCD is applicable. The first talk due to D. Kawell was a presentation of the spin programm at the PHENIX Collaboration, including new results on the gluon helicity distribution, from double helicity asymmetry measurements in $\pi^0$, direct photon or heavy flavors production. A short run at the highest energy $\sqrt{s}$=500GeV done in 2009 has given the first $W$ signals and the parity-violating asymmetry $A_L$ measurements, giving access to the quark and antiquark helicity distributions, are expected to be done in 2011. For tansverse spin phenomena, new results were obtained and in Fig.~6 one displays a sizeable single-spin asymmetry in forward $\pi^0$ production. More data are certainly needed for a full understanding of the rise in $x_F$ and the $p_T$ dependence of $A_N$.\\
A review talk on polarimetry at RHIC, for both PHENIX and STAR, was presented by A. Bazilevsky, who told us that $pp$ elastic scattering in the
Coulomb Nuclear Interference (CNI) region is ideal for absolute polarimetry in a wide energy range. He also recalled that a large forward neutron single-spin asymmetry, for $x_F>0$ was discovered in RHIC Run-2002, which is a very useful PHENIX local polarimetry. One can see in Fig.~7 the energy dependence of this asymmetry, very probably, a diffractive physics phenomena, whose theoretical interpretation is unfortunately not yet available.\\
Finally, we had two talks on STAR. First, L. Nogach presented new results on measurements of transverse spin effects in the forward region, for $\eta^0$ and $\pi^0$ production and discussed the future possibilities for similar measurements in jet production, $\Lambda$ production and Drell-Yan production of dilepton pairs. Second, Q. Xu discussed the results on the longitudinal spin transfer in $\Lambda$ and $\bar \Lambda$ inclusive production at $\sqrt{s}$=200GeV, which is well described by pQCD.

\begin{figure}[h]
\hspace{2.5pc}
\includegraphics[width=32pc]{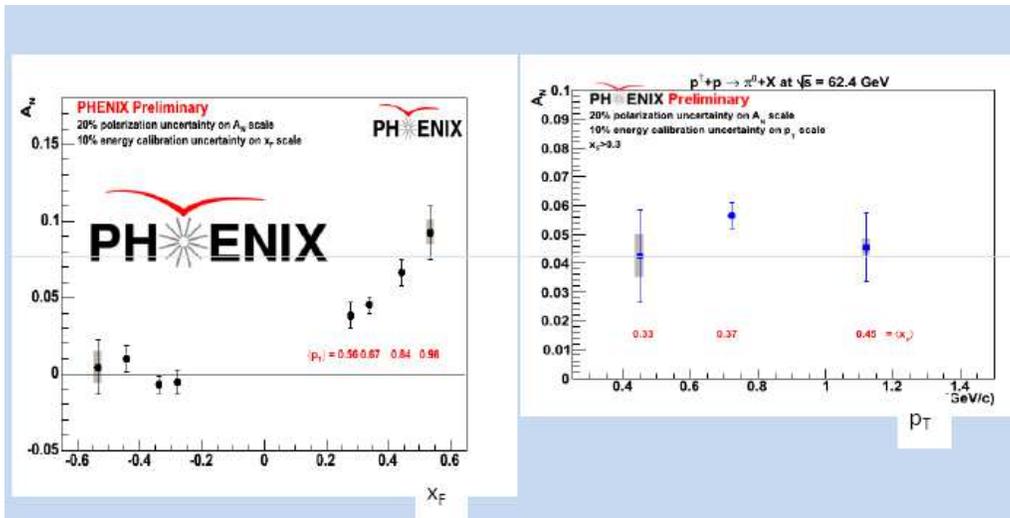}
\caption{\label{label}
Single transverse spin asymmetry for $\pi^0$ inclusive production at $\sqrt{s}=62.4\mbox{GeV}$ (Taken from Kawell's talk \cite{DK}).}
\end{figure}
\begin{figure}[h]
\hspace{3pc}
\includegraphics[width=30pc]{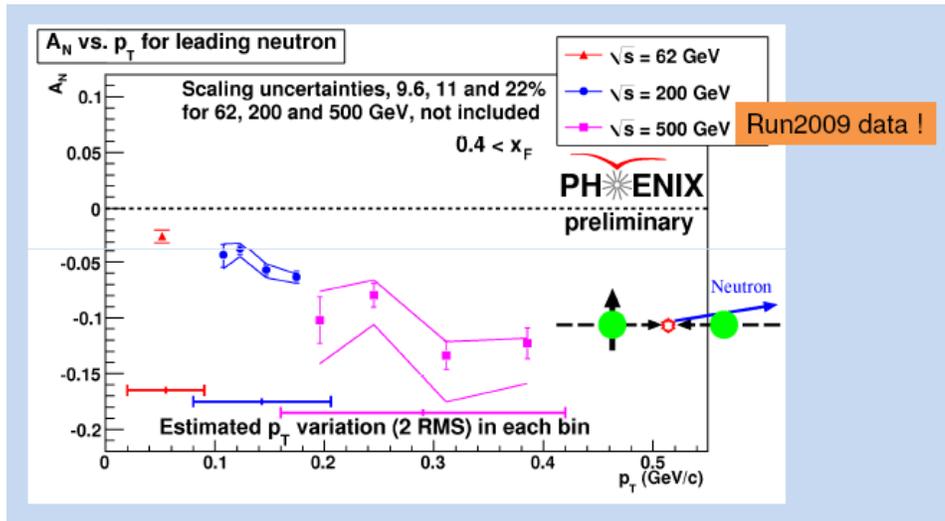}
\caption{\label{label}
Single transverse spin asymmetry for forward neutron inclusive production at $\sqrt{s}=62, 200, 500\mbox{GeV}$ (Taken from Bazilevsky's talk \cite{AB}).}
\end{figure}
\section{Theory}
Concerning the subject of phenomenological studies of parton distributions functions (PDF), we heard a talk on some new developments in the quantum statistical approach of the unpolarized and polarized PDFs, stressing several very challenging points, in particular in the high $x$ region \cite{JS}. In his presentation A. Sidorov discussed different methods of QCD analysis of the polarized DIS data. He emphasized the importance of higher twist effects and showed that the correct determination of the PDFs depends crucially whether or not they are taken into account. In another contribution O. Shevchenko presented a new NLO QCD parametrization of the polarized PDF, obtained by using all published results on DIS and SIDIS asymmetries. It
generalizes an earlier parametrization released by COMPASS, based on only inclusive DIS data. New SIDIS data from COMPASS, which will be available in the near future, should allow to improve the quality of this parametrization. Another method based on evolution equations for truncated Mellin moments of parton densities has been presented by D. Kotlorz. It avoids uncertainties related to some poorly known $x$ regions and may be a new tool for
an accurate reconstruction of the PDFs. A relevant question concerning the kinematic regions in $x$ and $Q^2$, where DGLAP is legitimately applicable, in connection with the infrared dependence of the structure function $g_1$, was discussed by B. Ermolaev. He recalled that the extrapolation of DGLAP
into the small-$x$ region is usually done by introducing in the parametrization of the PDFs, singularities which are theoretically groundless.\\
Ph. Ratcliffe first recalled us how painfull it was for the QCD community to accept the existence of large single-spin asymmetries, before new QCD
mechanisms were discovered. Then he discussed colour modification of factorization by relating the pQCD evolution of the Sivers function and the
twist-three gluonic pole contribution.\\
A model independent determination of fragmentation functions (FF) was proposed by E. Christova, by considering cross-section differences. It does not provide the full information on the FFs, only part of it, but without any assumptions on the FFs and on the PDFs.\\
Some aspects of quark motion inside the nucleon have been discussed by P. Zavada. The Cahn effect, which is an important tool to measure the quark transverse motion, and the unintegrated unpolarized PDFs $f(x,\bf{k_T})$, are studied in a covariant approach. In the framework of light-cone quark models, B. Pasquini presented a set of transverse momentum dependent (TMD) PDFs, allowing to derive some azimuthal spin asymmetries which were compared with available experimental data from CLAS, COMPASS and HERMES.\\
X. Artru described a very simple recursive fragmentation model with quark spin and its possible application to jet handedness and Collins effects for
pions, either emitted directly or coming from vector meson decay.\\
In his talk O. Teryaev discussed several issues related to the spin puzzle, the axial anomaly, the strangeness polarization and the role of heavy quarks polarization, suggesting a possible charm-strangeness universality and questioning when strange quarks can be heavy.\\
Cross sections and spin asymmetries in light vector meson leptoproduction were analyzed in the framework of generalized parton distributions by S.V. Goloskokov and the results are in good agreement with data from HERA, HERMES and COMPASS at various energies. Along the same lines, P. Kroll presented
a talk on spin effects in hard exclusive meson electroproduction within the framework of the handbag approach, showing the failure of the leading-twist
calculations.\\ 

{\bf Acknowledgments}\\

I am thankful to the organizers of DSPIN09 for their warm hospitality at JINR and for their invitation to present this summary talk. I am also
grateful to all the conference speakers for the high quality of their contributions. My special thanks go to Prof. A.V. Efremov for providing a full financial support and for making, once more, this meeting so successful.

\end{document}